\documentclass[useAMS,usenatbib]{mn2e}

\usepackage[latin1]{inputenc}
\usepackage{multirow}
\usepackage{graphicx}
\usepackage{multicol}
\usepackage{float}
\usepackage{footnote}
\usepackage{amssymb}
\usepackage[figuresright]{rotating}
\usepackage{amsmath}
\usepackage{url}
\usepackage{rotating}
\usepackage{caption}
\newcommand{\Mpc}{\rm\; Mpc}
\newcommand{\kpc}{\rm\; kpc}

\newcommand{\km}{\rm\; km}

\newcommand{\cm}{\rm\; cm}
\newcommand{\mum}{\hbox{$\rm\; \mu m\,$}}

\newcommand{\s}{\rm\; s}

\newcommand{\ks}{\rm\; ks}

\newcommand{\GHz}{\rm\; GHz}

\newcommand{\keV}{\rm\; keV}

\newcommand{\erg}{\rm\; erg}

\newcommand{\ergps}{\hbox{$\erg\s^{-1}\,$}}

\newcommand{\kmps}{\hbox{$\km\s^{-1}\,$}}

\newcommand{\kmpspMpc}{\hbox{$\kmps\Mpc^{-1}\,$}}

\newcommand{\Lx}{\hbox{$\thinspace L_\mathrm{X}$}}

\newcommand{\pcmsps}{\hbox{$\cm^{2}s^{-1}$}}

\newcommand{\Omm}{\hbox{$\rm\thinspace \Omega_{m}$}}
\newcommand{\OmL}{\hbox{$\rm\thinspace \Omega_{\Lambda}$}}

\title[The cluster around 4C+55.16]{AGN feedback and iron enrichment in the powerful radio galaxy, 4C+55.16}
\author[J. Hlavacek-Larrondo, et al.]{J. Hlavacek-Larrondo$^{1}$\thanks{E-mail: juliehl@ast.cam.ac.uk}, A. C. Fabian$^{1}$, J. S. Sanders$^1$ and G. B. Taylor$^{2}$\\
$^{1}$Institute of Astronomy, University of Cambridge, Madingley Road, Cambridge CB3 0HA\\
$^{2}$Department of Physics and Astronomy, University of New-Mexico, Albuquerque, NM 87131, USA}
\begin{document}

\date{Accepted 2011 April 26. Received 2011 April 22; in original form 2011 March 13}

\pagerange{\pageref{firstpage}--\pageref{lastpage}} \pubyear{2011}

\maketitle

\begin{abstract}
We present a detailed X-ray analysis of 4C+55.16, an unusual and interesting radio galaxy, located at the centre of a cool core cluster of galaxies. 4C+55.16 is X-ray bright ($\Lx{\scriptstyle\rm {(cluster)}}\sim10^{45}\ergps$), radio powerful, and shows clear signs of interaction with the surrounding intracluster medium. By combining deep $Chandra$ (100 ks) with 1.4 GHz VLA observations, we find evidence of multiple outbursts from the central AGN, providing enough energy to offset cooling of the ICM ($P_{\rm bubbles}=6.7\times10^{44}\ergps$).  Furthermore, 4C+55.16 has an unusual intracluster iron distribution showing a plume-like feature rich in Fe L emission that runs along one of the X-ray cavities. The excess of iron associated with the plume is around $10^7M_{\odot}$. The metal abundances are consistent with being Solar-like, indicating that both SNIa and SNII contribute to the enrichment. The plume and southern cavity form a region of cool metal-rich gas, and at the edge of this region, there is a clear discontinuity in temperature (from $kT\sim2.5\keV$ to $kT\sim5.0\keV$), metallicity (from $\sim0.4Z_{\odot}$ to $\sim0.8Z_{\odot}$), and surface brightness distribution, consistent with it being caused by a cold front. However, we also suggest that this discontinuity could be caused by cool metal-rich gas being uplifted from the central AGN along one of its X-ray cavities. 
\end{abstract}


\section{Introduction}

AGN feedback plays a major role in quenching star formation, enriching the surrounding medium with metals, and fuelling the supermassive black hole (SMBH) of the host galaxy. Major advancements in understanding the details of how AGN feedback operates have been possible through detailed studies of X-ray cavities \citep[see a review on the topic by][]{Mcn200745,Pet2006427}. The SMBH lying at the centre inflates these cavities, also known as bubbles, through jets of relativistic plasma. Bubbles therefore provide a direct measurement of the energy being injected by the SMBH into the surrounding medium.

In cool core clusters of galaxies, the active galactic nuclei (AGN) lying at the centre can energetically offset cooling of the intracluster medium by inflating these large cavities correlated with radio lobes, inducing weak shocks and propagating energy through sound/pressure waves \citep{Bir2004607,Raf2006652,Dun2006373,Dun2008385,Mcn200745,Fab2003344,Fab2006366,For2005635,San2007381}. 

This paper aims to study the AGN feedback processes arising in 4C+55.16, an unusual and interesting radio galaxy, located at the centre of a cool core cluster of galaxies at a redshift of $z=0.2412$ \citep{Pea1981248,Pea1988328,Why1985214,Hut198866,Sch2007134}. 4C+55.16 is X-ray bright ($\Lx\sim10^{45}\ergps$), radio powerful ($L_{\rm R}=8$Jy beam$^{-1}$ at 1.4GHz), and shows clear signs of interaction between its central galaxy and the intracluster medium (ICM) \citep[see][]{Why1985214,Iwa1999306,Iwa2001328}, providing an interesting case for studying AGN feedback in clusters of galaxies. 
\begin{figure*}
\centering
\begin{minipage}[c]{0.99\linewidth}
\centering \includegraphics[width=\linewidth]{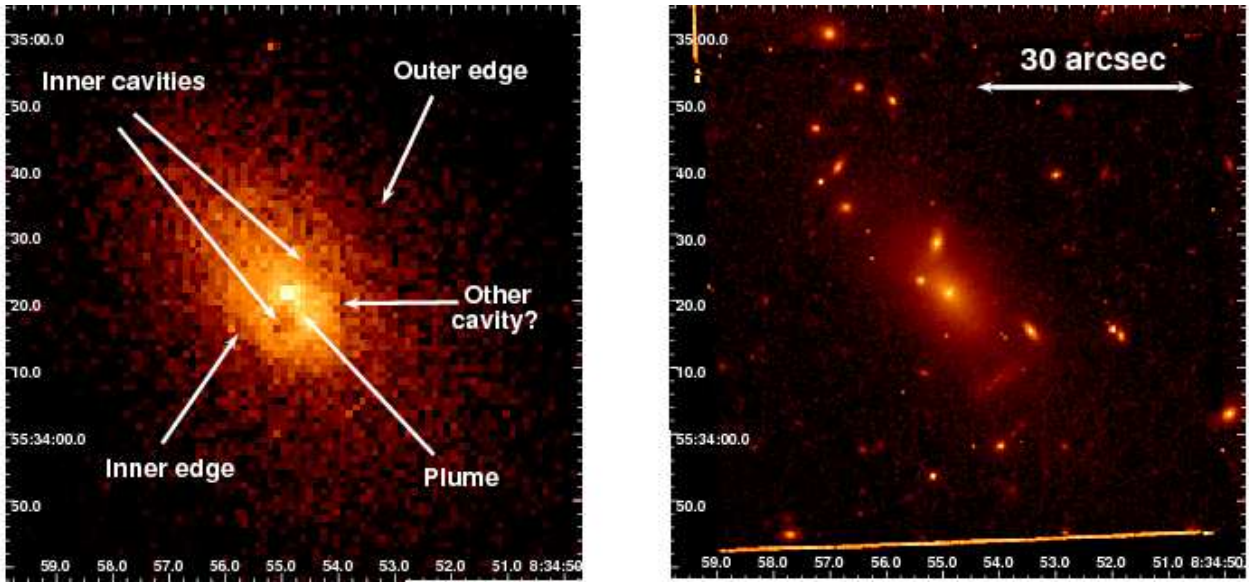}
\end{minipage}
\caption{Left - Exposure-corrected $0.5-7\keV$ $Chandra$ image of 4C+55.16, spatially smoothed with a $1''$ gaussian function and covering $\sim500\times500\kpc$ (or $130\times130''$). Shown in this image are the inner cavities, which are filled with $1.4\GHz$ radio emitting particles, the other possible cavity (with no GHz radio emission), the outer and inner edge as seen in the X-ray, and the plume-like structure associated with the abundance increase. Right - $Hubble$ Legacy Archive image of the cluster \citep{Edg2000}. }
\label{fig1}
\end{figure*}

Furthermore, 4C+55.16 has an unusual intracluster iron distribution. Using a 10 ks {\em Chandra} exposure, \citet{Iwa2001328} found that there was a large increase in metallicity at a radius of about 10 arcsec ($\sim40\kpc$), that went from half solar to twice solar. The authors suggested that this increase was due to a plume-like structure located in the south-west side of the cluster, which had a strong Fe L emission feature. The reason as to how so much iron could be accumulated remained unclear. Here, we use significantly deeper $Chandra$ observations of the source (100 ks) to study the abundance, and its distribution. By combining the $Chandra$ observations with VLA radio data, we make a detailed study of the X-ray cavities of the source. 

Section 2 presents the details of the data reduction. In Section 3 and Section 4, the techniques and results concerning the imaging and spectral analysis are respectively shown. Finally, we discuss the results in Section 5 and present the conclusions in Section 6. We adopt $H_\mathrm{0}=70\kmpspMpc$ with $\Omm=0.3$, $\OmL=0.7$ throughout this paper, and all error estimates are on the $1\sigma$ level. The redshift of the source corresponds to a scale of 3.807 kpc arcsec$^{-1}$. 
   
\section{Data reduction}
\subsection{$Chandra$ - X-ray}

4C+55.16 was originally observed with $Chandra$ on 2000 October 10 for 10 ks (ObsID 1645) with the $Chandra$ CCD Imaging Spectrometer (ACIS) in FAINT mode, such that the cluster was centred on the ACIS-S3 back-illuminated chip. It was subsequently observed on 2004 January 3 for $100$ ks (ObsID 4940) in VFAINT mode, significantly improving the image quality, and was centred on the ACIS-S3 back-illuminated chip, with the ACIS-S1, ACIS-S2, ACIS-I2 and ACIS-I3 also switched on. 

Both ObsID were processed, cleaned and calibrated using the latest version of the {\sc ciao} software ({\sc ciaov4.3}, {\sc caldb4.4.1}), and starting from the level 1 event file. We applied both CTI (charge time interval) and time-dependent gain corrections, as well as removed flares using the other back-illuminated chip (ACIS-S1) and the {\sc {lc$\_$clean}} script, with a $3\sigma$ threshold. 

For ObsID 4940, some strong background flares were seen towards the end of the exposure, and were removed. This resulted in a final exposure time of 73.6 ks. We then exposure-corrected the image, using an exposure map generated with a monoenergetic distribution of source photons at 1.5\keV  (which is almost the peak energy of 4C+55.16). The exposure-corrected $0.5-7.0\keV$ image for ObsID 4940 is shown on the left side of Fig. \ref{fig1}. We also reduced the previous observations (ObsID 1645), and although some fluctuations were seen in the data (about 20 percent from the mean in the light curve), we kept all data as did \cite{Iwa2001328}.

Spectra were analysed using {\sc xspec} \citep[{v12.6.0d,f, \rm e.g.} ][]{Arn1996101}. The Galactic absorptions throughout this paper were kept frozen at the \citet{Kal2005440} value,  $4.29\times10^{20}\cm^{-2}$ ($\mathrm{log}(N_\mathrm{H})=20.63$). Letting the absorption free to vary for the spectrum of the entire cluster did not improve significantly the fit. We also used the abundances of \citet{And198953} throughout this paper. 
\begin{figure*}
\centering
\begin{minipage}[c]{0.99\linewidth}
\centering \includegraphics[width=\linewidth]{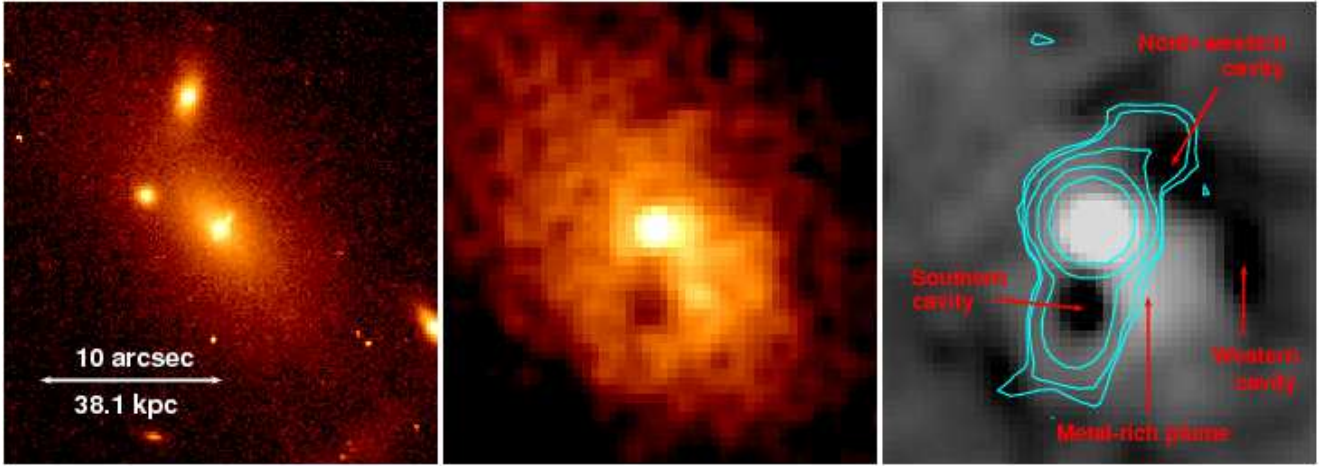}
\end{minipage}
\caption{Left: $Hubble$ Legacy Archive image of the central galaxy. This image shows evidence of an optical jet associated with the central point source. Middle: Smoothed $Chandra$ image of 4C+55.16 in the $0.5-7.0$ keV energy band, where the {\sc ciao} sub-pixel scripts have been applied. Right: Greyscale is an unsharp-masked $Chandra$ image where a 2D gaussian smoothed image ($\sigma=10$ pixels) was subtracted from a less smoothed image ($\sigma=2$ pixels), and which reveals deviations in the original $Chandra$ image. Also shown in light blue are the contours of VLA data at $1.4\GHz$ , configuration A (see Xu et al., 1995) with a beam size of $2.3''\times2.3''$. Contours are drawn between 0.003Jy beam$^{-1}$ to 0.403 Jy beam$^{-1}$ (5 per cent of the peak intensity), in a log scale. At least two X-ray cavities are seen: one to the south, and one to the north-west of the central point source. However, there seems to be another cavity to the west of the cluster. }
\label{fig2}
\end{figure*}
\subsection{VLA - Radio}
There exists a short exposure ($\sim 6$ mins) of published VLA observations at 1.4 GHz and bandwidth of 50 MHz in configuration A (Xu et al. 1995). This configuration is needed in using the VLA to produce maps with similar spatial resolution to $Chandra$. We retrieved the data from the {\sc NRAO} archive and reduced it in {\sc aips} ({\sc astronomical image processing system}) using the standard procedures. We used 3C286 as a flux calibrator, and the {\sc difmap} software to clean and self-calibrate the data \citep{She199426}.

The VLA observations reveal a strong core and twin lobes. The shorter of the two lobes points to the north-west, while the longer lobe is oriented almost directly to the south. 4C+55.16 has been imaged with VLBI \citep{Pol199598,Xu199599} and shows a core and one-sided jet pointing to the north-west. 

\section{Imaging analysis}
We use the deep $Chandra$ observations (100\ks) to analyse the physical state of the hot ICM, unless mentioned otherwise. Combining this data set with the 10 ks observations would not have improved significantly the signal-to-noise. 

In Fig. \ref{fig1} (right), we show the $Hubble$ Legacy Archive image of the cluster \citep{Edg2000}, and in Fig. \ref{fig2}, we show in same-scale images, the optical image taken from the $Hubble$ Legacy Archive, the $Chandra$ X-ray and unsharp-masked images, as well as the 1.4 GHz VLA radio contours of the central regions of the cluster. The $Chandra$ image was obtained after applying the {\sc ciao} sub-pixel scripts in order to obtain images with the best resolution possible. The unsharp-masked image was obtained by subtracting a strongly 2D gaussian smoothed image ($\sigma=10$ pixels or $\sigma=4.9''$) from a less smoothed image ($\sigma=2$ pixels), and reveals deviations in the original $Chandra$ image. At least two X-ray cavities are seen: one to the south, and one to the north-west of the central point source. However, there seems to be another cavity to the west of the cluster. The contours show that two radio lobes fill the X-ray cavities, consistent with the idea that these cavities are created by the central AGN. However, deeper observations would be needed to determine whether there is radio emission associated with the western cavity.  

In the optical, we can see the central galaxy surrounded by a diffuse envelope, along with a bright point source that coincides with the X-ray and radio point sources. We also see a small jet in the optical, where the axis of the jet is aligned with the radio filled X-ray cavities. There could be a counterpart of the jet in the X-rays, but the resolution is not good enough to resolve it (even with the {\sc ciao} sub-pixel scripts that give the best resolution for $Chandra$, see middle panel of Fig. \ref{fig2}).  
\begin{figure}
\centering
\begin{minipage}[c]{0.99\linewidth}
\centering \includegraphics[width=\linewidth]{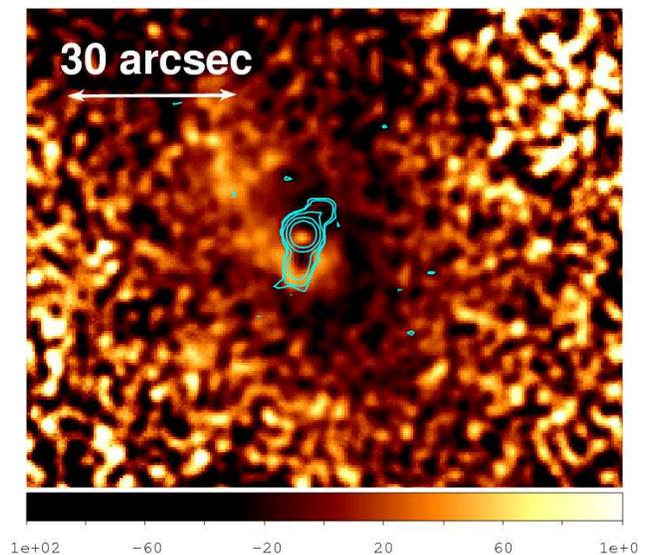}
\end{minipage}
\caption{Fractional difference between each pixel and the average of an ellipse centred on the X-ray point source and passing through the pixel. When constructing the ellipses, we use an average ratio between the major ($a$) and minor ($b$) axis of the ellipses of $a/b\sim1.6$, and an average rotated angle of the ellipses with respect to the north of $\theta\sim35^o$, counter-rotated. The image has then been smoothed with a $\sigma=2$ pixels Gaussian. Also shown in light blue are the contours of VLA data at $1.4\GHz$. }
\label{fig3}
\end{figure}

To further enhance the surface brightness fluctuations in the X-ray image, at each point, we subtract from the X-ray image the average value within an ellipse centred on the X-ray point source. We used an ellipse rather than a circle, since the surface brightness contours better follow this shape. Using the surface brightness contours, we determine the average ratio between the major ($a$) and minor ($b$) axis of the ellipses, as well as the average rotated angle of the contours with respect to the north ($\theta$, counter-clockwise). We find $a/b\sim1.6$ and $\theta\sim35^o$. Using these values, at each point in the X-ray image, we build an ellipse passing through the point and centred on the central X-ray point source, and then subtract the average value within the ellipse and calculate the fractional difference. We then smooth the image with a 2D gaussian function of $\sigma=2$ pixels. This gives the image shown in Fig. \ref{fig3}. Initially, we remove all point sources, except for the one associated with the central galaxy. To do this, we identify them with the {\sc ciao} program {\sc wavdetect}, and replace them with the average value of a surrounding background.

Fig. \ref{fig3} shows interesting structures, including bright rims surrounding the southern cavity, and a tail of enhanced material starting out from the southern cavity and plume-like feature, and making its way from east to north. 

\begin{figure}
\begin{minipage}[c]{0.7\linewidth}
\hspace{0.5in}
\includegraphics[width=\linewidth]{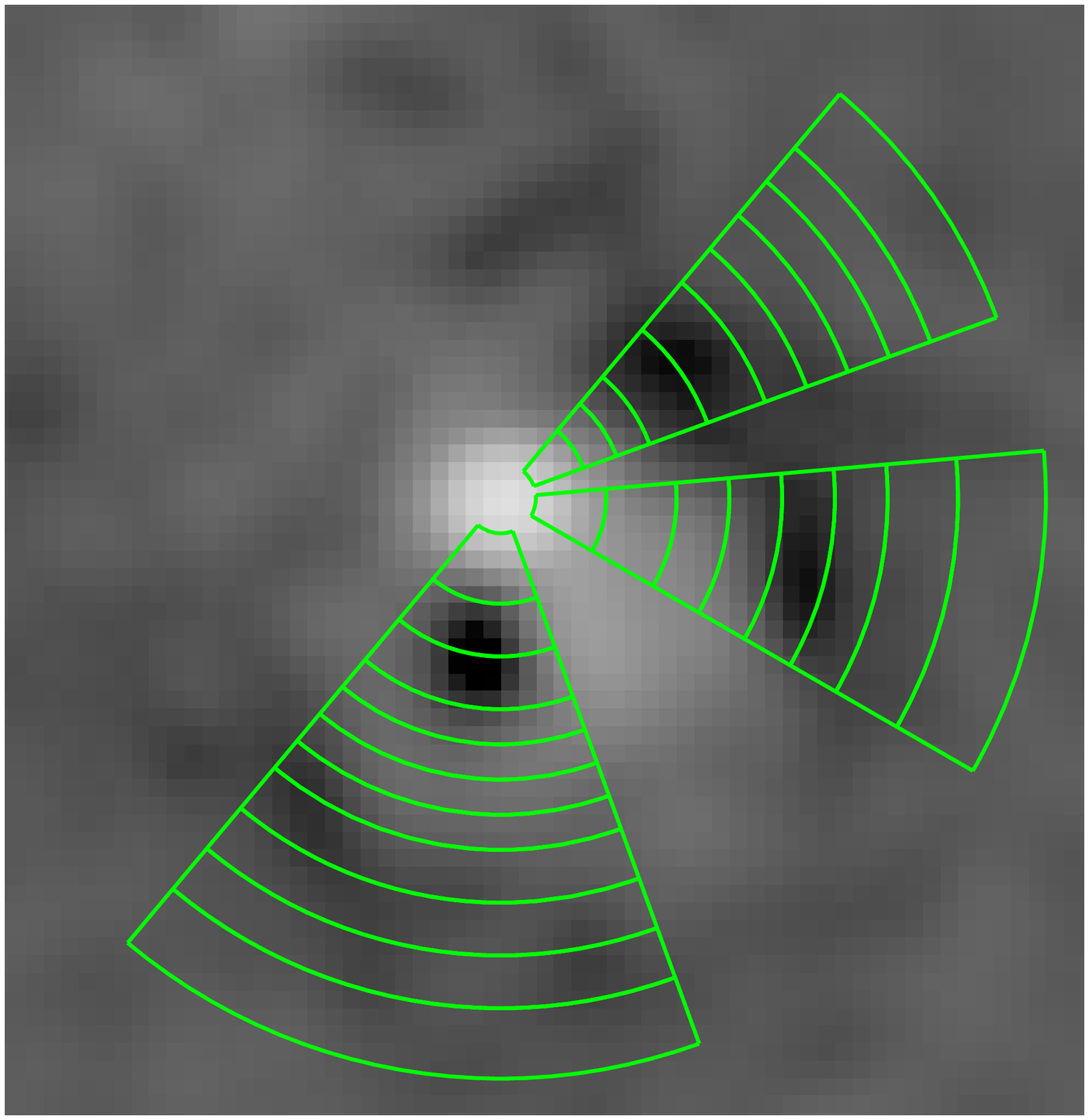}
\end{minipage}
\begin{minipage}[c]{0.99\linewidth}

\hspace{-0.2in}\includegraphics[width=\linewidth]{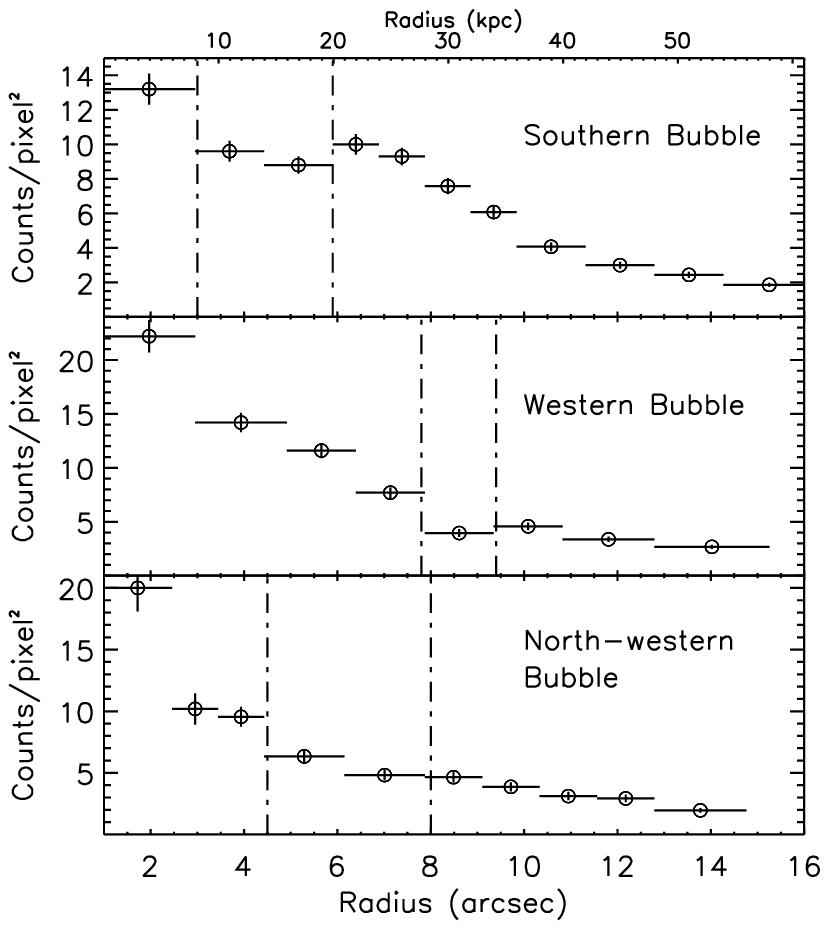}
\end{minipage}
\caption{Top: Unsharp-masked image shown in Fig. 1, with contours of regions used to look at the surface brightness distribution along each cavity. Bottom: Surface brightness distribution as a function of radius along each cavity. Also shown, are the locations of each cavity (between the dot-dashed lines). }
\label{fig4}
\end{figure}

As shown in the unsharp-masked image of Fig. \ref{fig1} (right), there are at least two X-ray cavities (one to the south, and one to the north-west). However, if the potential cavity to the west is real, we should see a disturbance in the surface brightness distribution. In the top panel of Fig. \ref{fig4}, we show different regions we selected along each of the cavities. For each region, we determined the average count-rate pixel$^{-2}$, and then plotted its distribution for each cavity as a function of radius in the lower panel of Fig. \ref{fig4}. This figure shows decrements in X-ray emission for each of the cavities, although it is quite difficult to tell for the north-western cavity. This cavity is however filled with radio emission, and we therefore still consider it as a cavity.

\section{Spectral analysis}

\subsection{Central AGN}
Using the deep $Chandra$ observations, we first examine the nucleus of 4C+55.16. Fig. \ref{fig2} shows a bright X-ray point source located at the centre of the cluster, which coincides with the radio point source and optical point source of the central galaxy. Although the X-ray point source is bright and point-like, we estimate that there is no significant pileup. Pileup occurs when two or more photons are detected as one event \citep[see for more details][]{Dav2001562,Rus2009402}. The significance of pileup can be examined by comparing the amount of good grades (grades 0,2,3,4,6) to the bad grades (grades 1,5,7) for our point-like nucleus. We find the ratio of bad/good to be $\sim0.03$. Typically, pileup starts becoming problematic when the fraction of bad grades exceeds  10 per cent of the good grades. 

To estimate the nuclear luminosity, we analyse the $0.5-7\keV$ spectrum of the inner $r=1''$ region, centred on the central point source. We use a surrounding annulus ($r=2-3''$) as a background, and bin the spectra with a minimum of 30 counts per bin. We fit an absorbed power-law model to the background-subtracted spectrum, and include {\sc phabs} absorption to account for Galactic absorption. Initially, we only consider Galactic absorption and keep it frozen to the \citet{Kal2005440} value. We use $\chi^2$-statistics to find the best fitting model, and let the photon index and normalisation free to vary. We find a photon index of $\Gamma=1.55\pm{0.17}$, and a reduced $\chi^2$ of 1.3. Using the {\sc cflux} model in {\sc xspec}, we obtain a flux estimate for the nucleus corrected for absorption, and then convert it into a luminosity using the luminosity distance. We find a $2-10\keV$ unabsorbed luminosity for the nucleus of $(1.1\pm{0.2})\times10^{43}\ergps$, and that the value does not depend significantly on the photon index to within $\pm0.3$. There is also no evidence suggesting that the nuclear luminosity has changed within the 3 year difference separating the observing dates of the 100ks and 10ks data.

If the Galactic absorption is also free to vary, we find that the model is no longer well constrained, and the estimated value for the absorption is not consistent with the values predicted by \citet{Kal2005440} . Adding an additional absorption at the redshift of the source does not improve the fit, and converges towards a null value for the additional absorption. We also fit a single-temperature {\sc mekal} \citep{Mewe1995} model to the nuclear spectrum, and find a reduced $\chi^2$ of 1.5. The parameters are not well constrained and the estimated temperature (${\rm kT}=10^{+10}_{-3}$ keV) is quite large compared to the surrounding gas.  

If the X-ray surface brightness profile of the cluster scaled with radius as a power-law, then it would predict a count rate of 30 counts pixel$^{-2}$ for the nucleus. However, the X-ray point source is about three times as bright (90 counts pixel$^{-2}$). Yet, its intrinsic luminosity remains $\sim100$ times fainter than the total X-ray luminosity of the cluster ($\Lx\sim10^{45}\ergps$; see Hlavacek-Larrondo \& Fabian 2011 for a discussion of this issue)\nocite{Hla2011}. As seen in Fig. \ref{fig2}, the optical shows evidence of a jet associated with the central point source. Since we are not able to resolve the jet in the X-rays, the nuclear luminosity we derive might have a beamed contribution from the jet, making the intrinsic AGN luminosity even fainter. 

\subsection{Temperature and abundance maps}
To study the temperature and abundance distribution across the cluster, we bin different regions together using a Contour Binning algorithm which follows the surface brightness variations \citep[see][]{San2006371}. This technique is better suited to analyse the temperature and abundance variations along radial structures, as opposed to azimuthal structures.

\begin{figure}
\centering
\begin{minipage}[c]{0.99\linewidth}
\centering \includegraphics[width=\linewidth]{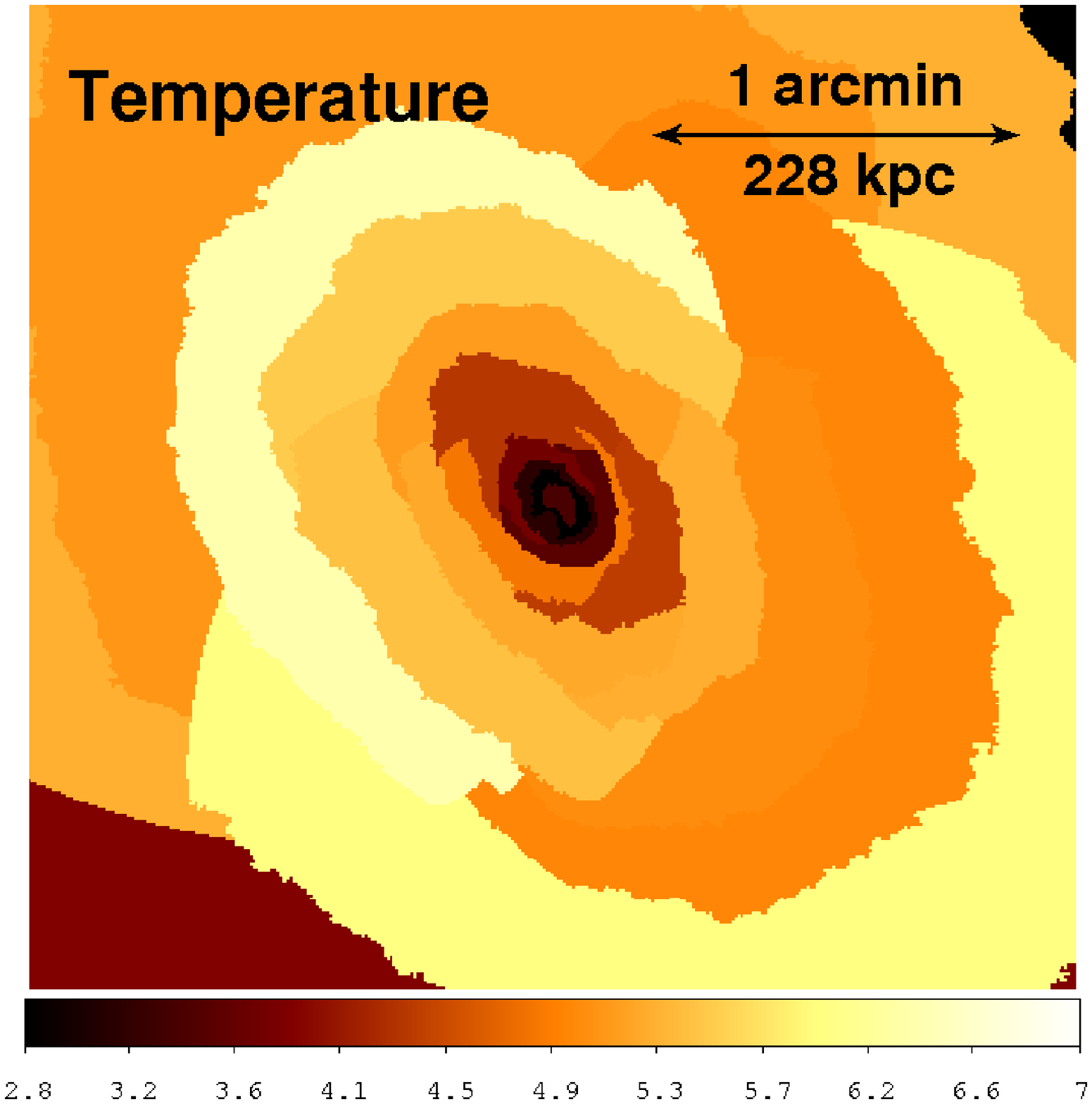}
\end{minipage}
\begin{minipage}[c]{0.99\linewidth}
\centering \includegraphics[width=\linewidth]{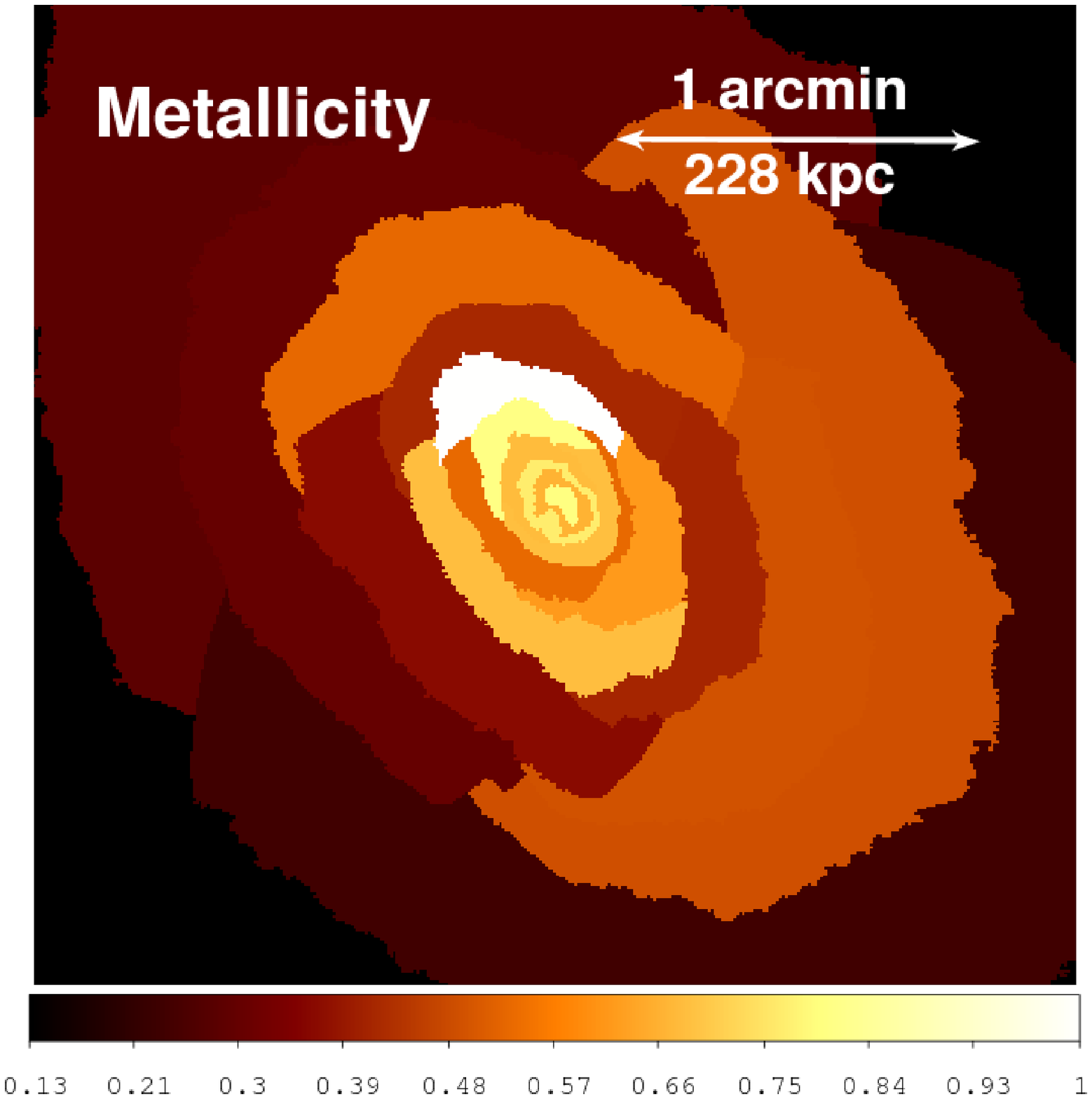}
\end{minipage}
\caption{Large scale temperature (top) and abundance (bottom) map of the cluster, as derived using the Contour Binning algorithm (see Section 4.2). We binned regions within $~900\times900$ kpc, with a signal-to-noise of 50. The temperature units are keV and those of the metallicity are Z$_\odot$.   }
\label{fig5}
\end{figure}

We begin by analysing the large scale structures within $~900\times900$ kpc. We bin the regions so that they have a signal-to-noise ratio of 50 ($2500$ counts), and restrict the lengths to be at most twice the widths. We extract a $0.5-7\keV$ spectrum for each region, and use a region further away, but still within the chip, as a background. We then fit an absorbed {\sc mekal} model to the data, and use C-statistics. We let the temperature, absorption and normalisation parameters free to vary. The resulting temperature and metallicity maps are shown in Fig. \ref{fig5}. For the temperature, the errors on each value vary from $\sim6$ per cent in the inner regions to $\sim30$ per cent in the outer regions, and for the abundance they vary from $\sim20$ per cent (inner) to $\sim40$ per cent (outer). 

The large-scale temperature map shows some structure, especially a temperature jump between the cluster core (inner $\sim50$ kpc) and surrounding gas on the southern side. The metallicity map also shows some structure. To the north-east, we see a first metal rich region (1$Z_\odot$) located at about 100 kpc, and then a second (0.6$Z_\odot$) at about 160 kpc. Further away ($r>200$ kpc), it seems that the north-east side is more metal poor than the south-western side. 

\begin{figure*}
\centering
\begin{minipage}[c]{0.33\linewidth}
\centering \includegraphics[width=\linewidth]{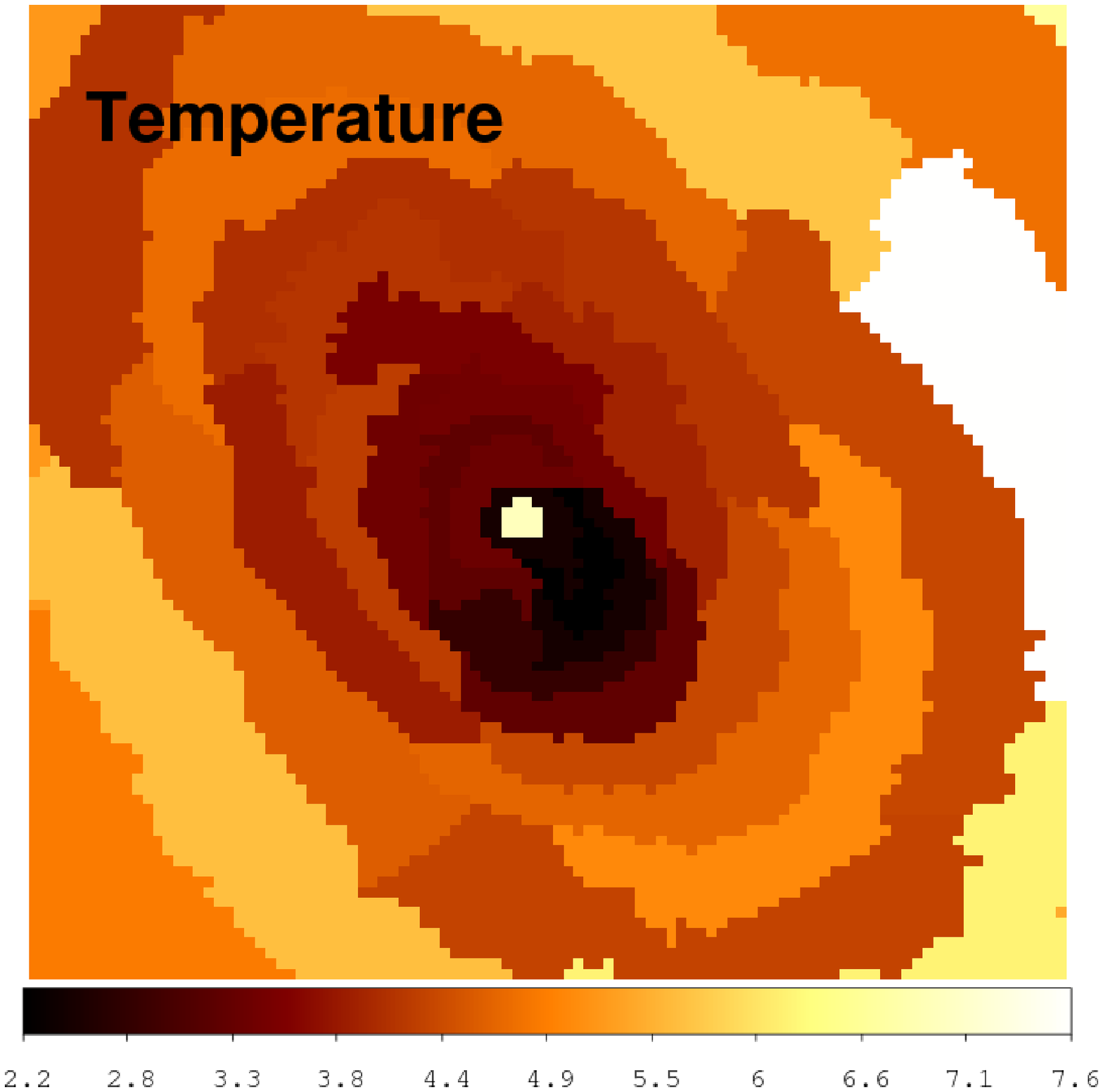}
\end{minipage}
\begin{minipage}[c]{0.33\linewidth}
\centering \includegraphics[width=\linewidth]{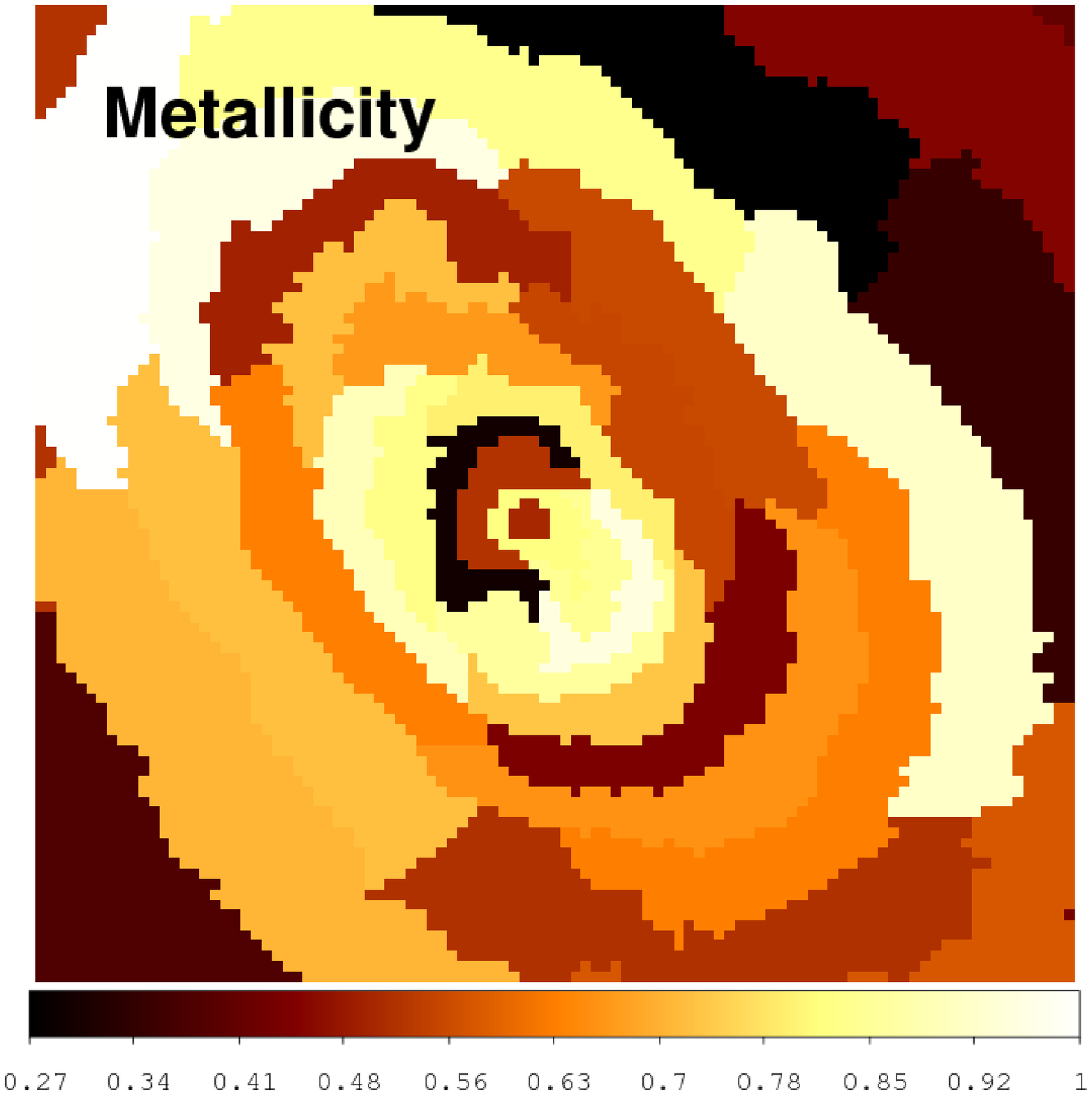}
\end{minipage}
\begin{minipage}[c]{0.315\linewidth}
\centering \includegraphics[width=\linewidth]{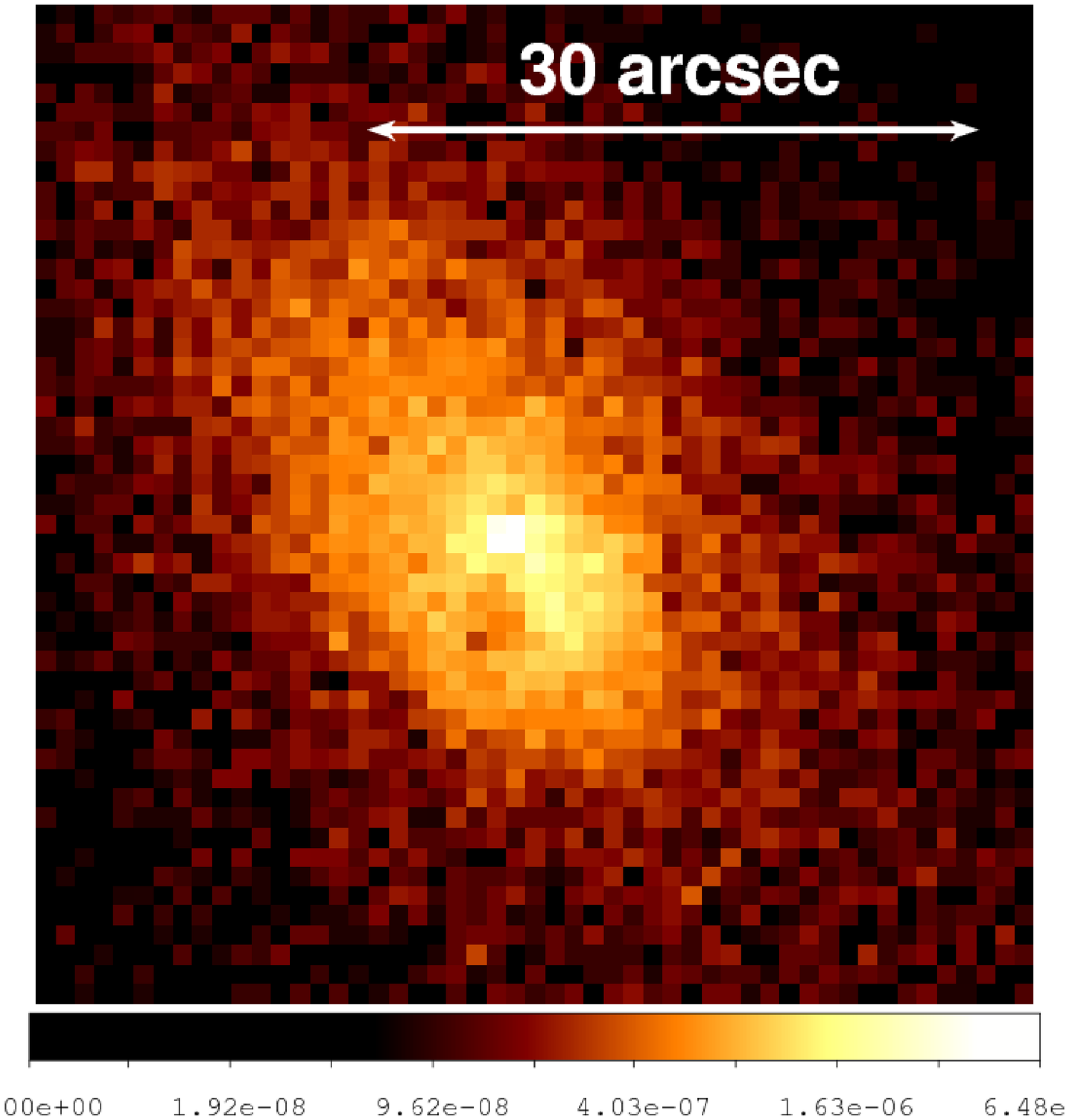}
\end{minipage}
\caption{Temperature (left) and abundance (middle) maps of the inner regions, as derived using the Contour Binning algorithm (see Section 4.2), as well as exposure-corrected image of the inner regions of the cluster. We binned regions within $~190\times190$ kpc, with a signal-to-noise of 30. The temperature units are keV and those of the metallicity are Z$_\odot$.}
\label{fig6}
\end{figure*}

Next, we analyse the inner regions of the cluster, within $~190\times190$ kpc. We bin the regions so that they have a signal-to-noise ratio of 30 (~900 counts), and also use C-statistics. The resulting temperature and abundance maps are shown in Fig. \ref{fig6}. Here, the errors on each value vary from $\sim9$ per cent in the inner regions to $\sim20$ per cent in the outer regions for the temperature, and from $\sim30$ per cent (inner) to $\sim60$ per cent (outer) for the abundance. 

The temperature map now reveals more clearly the temperature jump seen between the core and surrounding region just below the southern cavity and plume-like feature. The temperature jumps by a factor of two, from colder to warmer, and is similar to those seen in cold fronts. In the temperature map, we see a second jump in temperature to the west of the cluster at about $r\sim120$ kpc. This time, the temperature jumps from a colder ($kT\sim4$ keV) to warmer region ($kT\sim8$ keV) (see also Section 4.3.2). The metallicity map also clearly reveals that the plume-like feature (located to the south-west at $r\sim1-10''$) is metal rich when compared to a region within the same radius but excluding the plume (see also Section 4.3.4).

\subsection{Spectral profiles}

\subsubsection{Previous $Chandra$ observations (10ks)}
Using the 10 ks {\em Chandra} observations, \citet{Iwa2001328} found that there was an large increase in metallicity at a radius of about 10 arcsec ($\sim40\kpc$), that went from half solar to twice solar, which they associated with the plume-like feature. 

In order to confirm this increase in metallicity, we reduce the 10ks observations using the latest $Chandra$ reduction scripts. We then proceed to analyse the data in the same way as in \citet{Iwa2001328}, but choose to use the value derived by \citet{Kal2005440} for the Galactic absorption ($4.29\times10^{20}\cm^{-2}$) instead of the one derived by \citet[][ $4.20\times10^{20}\cm^{-2}$]{Dic199028}. Since these values are almost the same, we do not expect our results to vary significantly based on this choice of parameter.

\citet{Iwa2001328} argued that the redshift inferred from optical spectroscopy was not consistent with the one inferred by the Fe K line. From their analysis of the Fe K line, they derived a redshift of $z\sim0.254^{+0.010}_{-0.009}$. However, our spectra indicates otherwise. Using the deep 100 ks data, we extract a $0.5-7\keV$ spectrum within a radius of $r=1-9''$, since this region is known to contain a strong Fe K line (see also Fig. \ref{fig8}). We then fit an absorbed {\sc mekal} model to the spectrum, and let the redshift, temperature, abundance and normalisation parameter free to vary. We use $\chi^2$ statistics, and find a redshift of $z=0.238\pm{0.007}$, consistent with the value inferred by optical spectroscopy, and therefore choose to use the optically derived redshift ($z=0.2412$) throughout this paper.  

\begin{figure}
\centering
\begin{minipage}[c]{0.99\linewidth}
\centering \includegraphics[width=\linewidth]{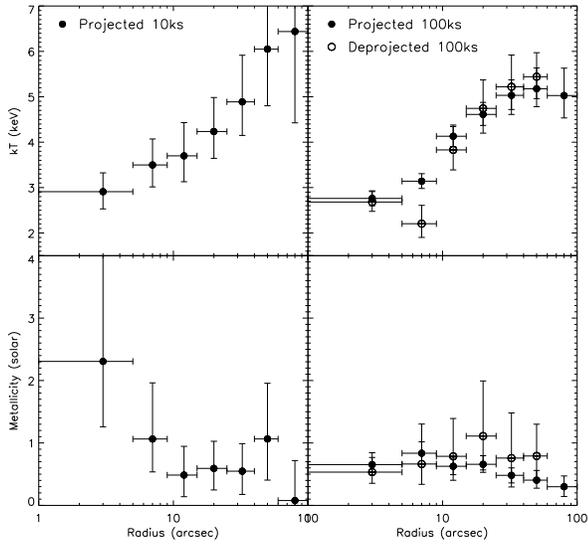}
\end{minipage}
\caption{Temperature (top) and metallicity (bottom) profiles for the regions as selected in \citet{Iwa2001328} ($r=1-2.5,2.5-5,5-9,9-15,15-25,25-40,40-60''$). The left panels show the results for the 10 ks observations, reduced using the latest $Chandra$ calibrations, and those on the right show the results derived from the 100 ks observations (projected and deprojected). The deeper data do not show the increase in metallicity seen in the 10 ks observations and reported in \citet{Iwa2001328}.}
\label{fig7}
\end{figure}

\begin{figure}
\centering
\begin{minipage}[c]{0.99\linewidth}
\centering \includegraphics[width=\linewidth]{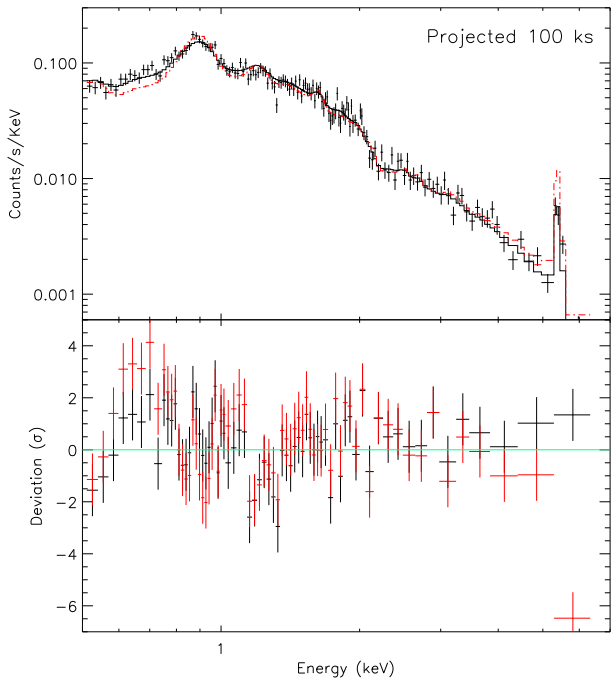}
\end{minipage}
\caption{Spectrum of the central $r=1-9''$ region (excluding the nucleus). The black curve shows the best-fitting model obtained for an absorbed (Galactic) {\sc mekal} model ($kT=2.93\pm{0.11}$\keV,  $Z=0.74\pm{0.12}$ solar). The red curve shows the best-fitting model while keeping the temperature and abundance frozen to the values found by \citet{Iwa2001328} for the same region ($kT=3.34\pm{0.35}$\keV, $Z=1.93^{+0.85}_{-0.62}$ solar).}
\label{fig8}
\end{figure}

We now proceed to take spectra from seven annuli, following the spectral profiles extracted in \citet{Iwa2001328} ($r=1-2.5,2.5-5,5-9,9-15,15-25,25-40,40-60''$)\footnote[1]{\citet{Iwa2001328} used eight profiles instead of seven, splitting the first bin into two bins of $r=1-2.5''$ and $r=2.5-5''$. However, our fit could not converge for each of these bins separately, and we were forced to merge them together in order to get a good enough S/N ratio and derive parameters.}. These annuli exclude the nucleus. We then fit an absorbed {\sc mekal} model to each spectrum, and derive the temperature, abundance and normalisation parameter using $\chi^2$ statistics. The results are shown in the left panels of Fig. \ref{fig7}. We do the same using the deep $Chandra$ 100 ks observations, but since the data is much deeper, we also include deprojection effects. The right panels of Fig. \ref{fig7} show the results we obtain for both the projected and deprojected data. The latter are obtained following the Direct Spectral Deprojection method of \citet{San2007381}.  

As shown in Fig. \ref{fig7}, the 100 ks observations give much tighter constraints on the temperature and abundance profiles, and most importantly they do not show the abundance increase seen in the 10 ks observations. The abundance increase we derive from the 10 ks observations is also not the same as in \citet{Iwa2001328}, who found an increase that went from half solar to twice solar. We only find an increase form half solar to one solar. However, when analysing in more detail the 10 ks observations, we first notice that the model is not very sensitive to the abundance. Even if we freeze the abundance to four times solar, it only increases the $\chi^2$ from 40.15 to 42.19. The abundance increase we see from the 10 ks observations should therefore be taken lightly. 

Secondly, we analyse the region between $r=1-9''$, as did \citet{Iwa2001328}, but using the deep 100 ks observations. We fit an absorbed {\sc mekal} model to the spectrum, and derive a best fitting value of $kT=2.93\pm{0.11}$ keV and $Z=0.74\pm{0.12}$ solar. When analysing this region, \citet{Iwa2001328} found using the 10 ks observations that $kT=3.34\pm{0.35}$ keV and $Z=1.93^{+0.85}_{-0.62}$ solar. In Fig.\ref{fig8}, we plot our spectrum, as derived from the 100 ks observations. Overlaid on top, we plot our best fitting model (black curve), and a model where we kept $kT$ and $Z$ frozen to the values derived by \citet{Iwa2001328}, and only allowed the normalisation parameter free to vary. As shown in the figure, the parameters derived by \citet{Iwa2001328} seem to overestimate the strength of the Fe K line, and also underestimate the soft energy part of the spectrum. We suspect the latter may be caused by the new calibration for the absorption model of the detector \citep[see for more details ][]{Mar20045165}, which now better corrects for the ACIS $Chandra$ contaminant at soft energies. Hence, it is possible that the earlier reduction scripts over corrected the spectrum at soft X-rays, and allowed the fit to converge toward higher abundances. 

\begin{figure}
\centering
\begin{minipage}[c]{0.99\linewidth}
\centering \includegraphics[width=\linewidth]{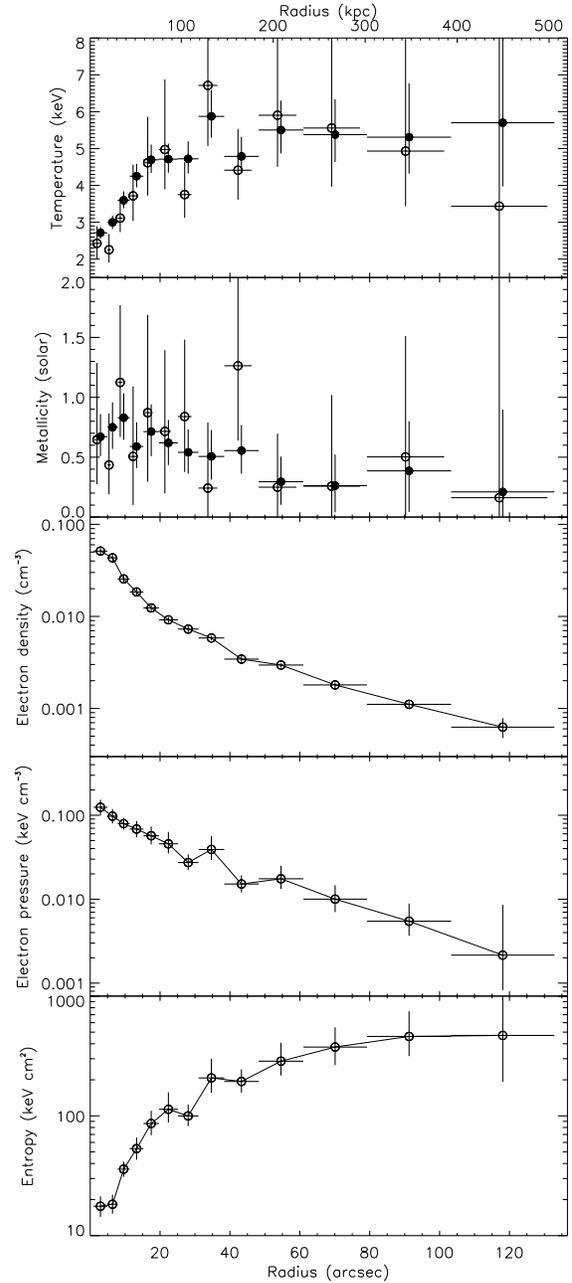}
\end{minipage}
\caption{Projected (filled symbols) and deprojected (non-filled symbols) temperature and metallicity profiles, as well as deprojected electron density, electron pressure and entropy profiles of the cluster, as derived by selecting annuli with a minimal signal-to-noise ratio of 63 ($\sim4000$ counts).  }
\label{fig9}
\end{figure}

\subsubsection{Deep $Chandra$ observations (100ks)}
We now proceed to do a more in depth analysis of the 100 ks observations. The regions selected by \citet{Iwa2001328} each had a different number of counts, ranging from $\sim2000$ counts to $\sim10000$ counts. Instead, we choose to select regions containing roughly the same signal-to-noise (S/N$\sim63$ or $\sim4000$ counts), and push the analysis to larger radii. We selected a background region within the same chip (ACIS-S3) but far from the cluster. The net count rate per pixel$^2$ of the selected background region is also the same as the count rate per pixel$^2$ of the ACIS-S1 chip (which is also back-illuminated), and should therefore not contain much cluster emission.

\begin{figure}
\begin{minipage}[c]{0.65\linewidth}
\hspace{0.75in}
\includegraphics[width=\linewidth]{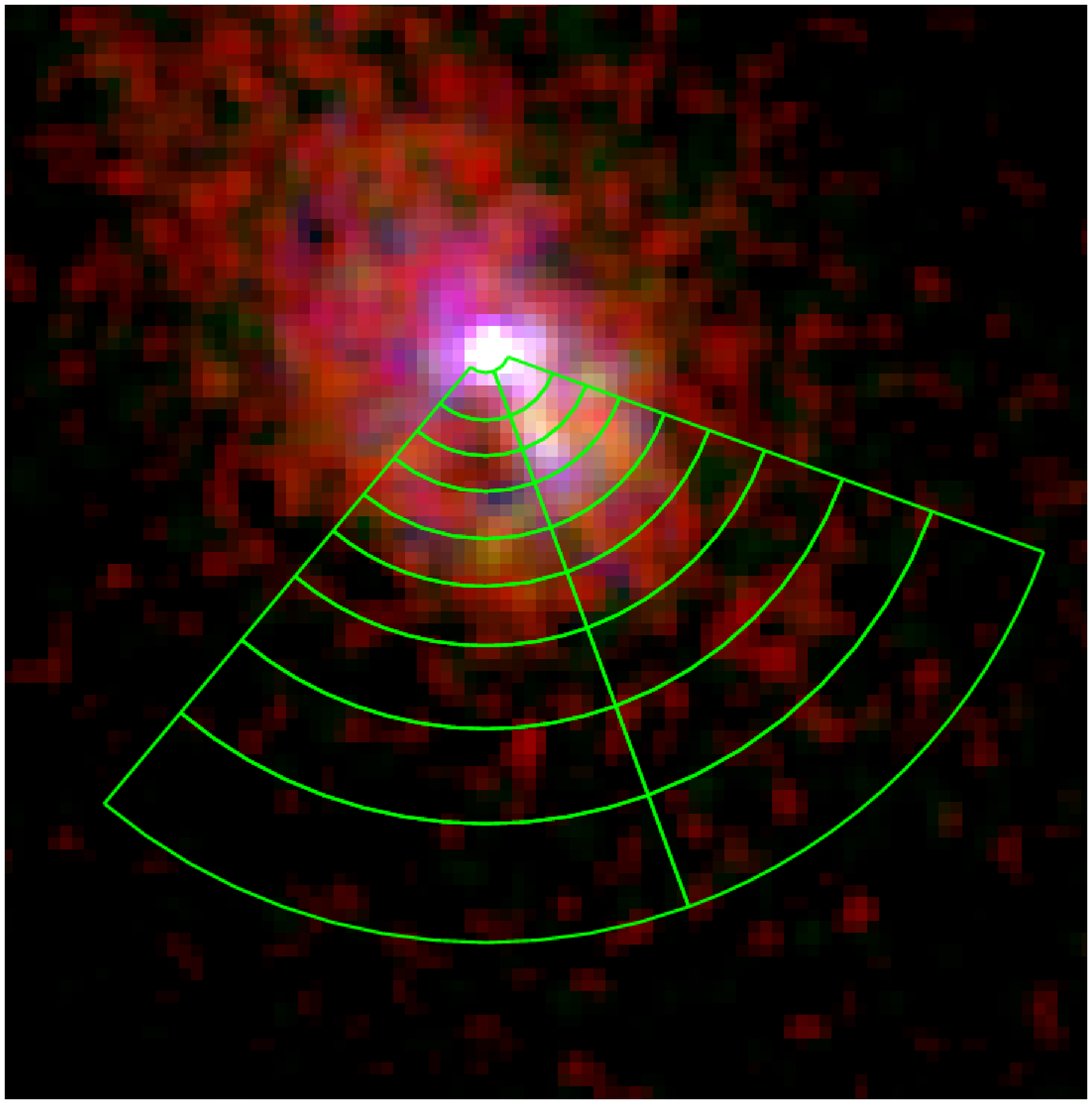}
\end{minipage}
\begin{minipage}[c]{0.98\linewidth}
\centering \includegraphics[width=\linewidth]{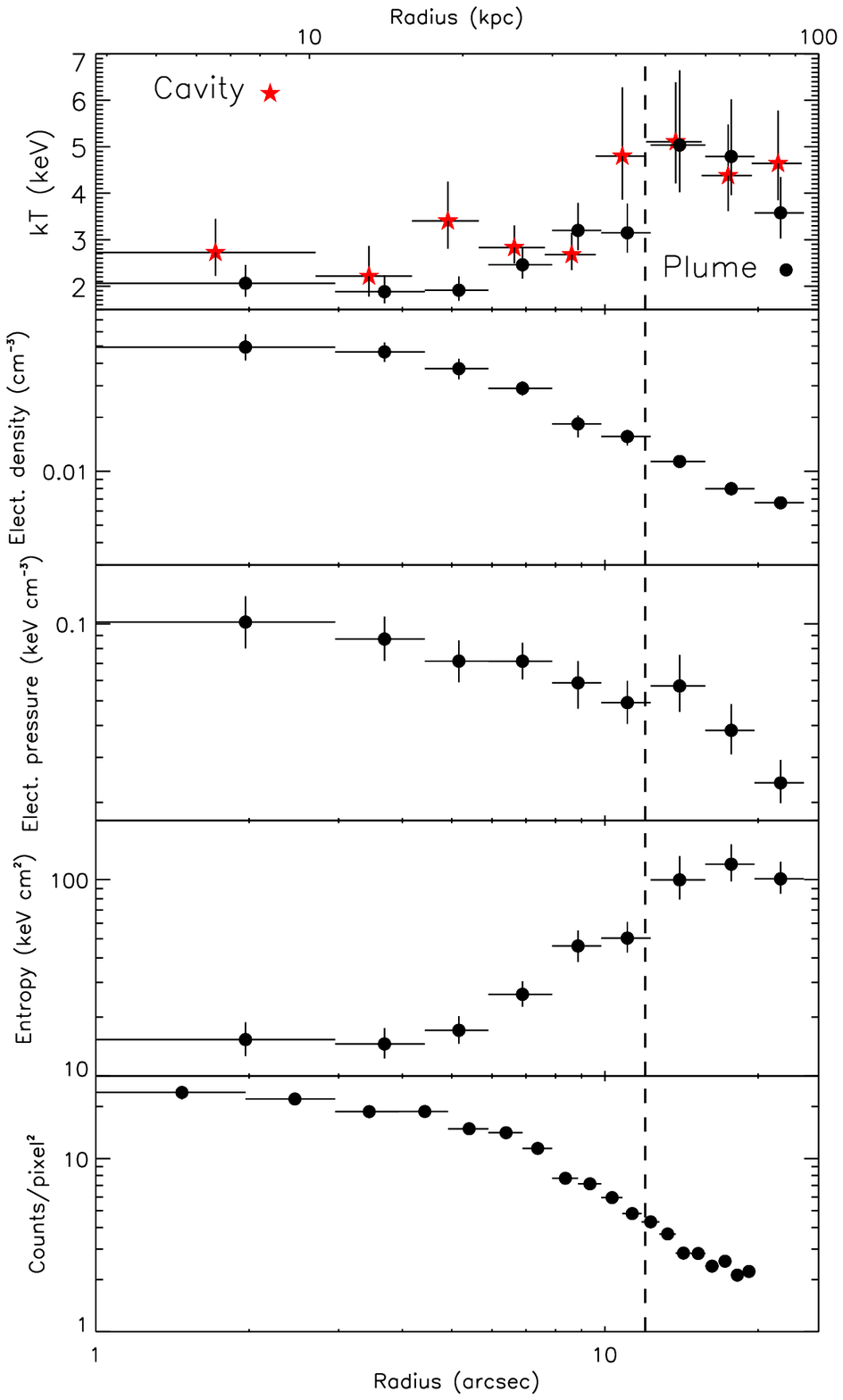}
\end{minipage}
\caption{Projected temperature profiles along the southern cavity (red) and plume-like feature (black), derived using a single {\sc mekal} model, as well as electron density, electron pressure and entropy profiles of the plume-like feature. The potential cold front is located at $r\sim40\kpc$. Selected regions are shown in the top figure which is a 3 colour image (red $0.3-0.7$, green $0.7-1.0$, and blue $1.0-7.0$ keV). The metal-rich plume stands out in green as it is dominated by the Fe L emission feature, centred at $\sim0.8$ keV.}
\label{fig10}
\end{figure}

The projected and deprojected temperature and metallicity, as well as deprojected electron density, electron pressure and entropy profiles ($S=kTn_{\rm e}^{-2/3}$) as a function of radius are show in Fig. \ref{fig9}. For each spectrum, we fitted an absorbed {\sc mekal} model, and used $\chi^2$ statistics. Although we include in Fig. \ref{fig9} the deprojected results, we found that they were consistent with the projected ones to within $1\sigma$ error estimates. Two-temperature models were also not needed, as they did not improve the fit. 

Fig. \ref{fig9} shows interesting features at $r>100$ kpc. The temperature structure remains constant at large radii, at least within the error measurements, while the metallicity structure on average decreases with radius (from $Z\sim0.75$ solar to $Z\sim0.2$ solar). 

There seems to be a slight increase in metallicity at $r\sim160$ kpc. This could be due to the same feature noticed earlier while analysing the large-scale metallicity map (see bottom panel of Fig. \ref{fig5}), which showed a slight increase at a similar radius on the north-eastern side of the cluster. However, the increase in metallicity is well within the $1\sigma$ uncertainties, and therefore does not appear to be statistically significant. If we compare the spectra of the different radial intervals with one another, we find no significant change in the iron line complexes.  

However, there appears to be an increase in temperature at $r\sim130$ kpc (Fig. \ref{fig9}), accompanied by a break in electron density, electron pressure and entropy, to more than a $1\sigma$ level. This could be associated with the warmer region noticed earlier in the temperature map of the inner regions (Fig. 6), where we could see a jump in temperature from colder ($kT\sim4$ keV) to warmer ($kT\sim8$ keV). 

\subsubsection{southern X-ray cavity}
We now concentrate on the spectral analysis along the southern cavity. Clear bright rims can be seen surrounding this cavity, and could indicate the presence of a shock front (see Fig. \ref{fig3}). Using regions along the southern cavity, we extract a spectrum, and fit an absorbed {\sc mekal} model to each region. The absorption is again taken as the purely Galactic. The resulting temperature profile is shown in Fig. \ref{fig10}. This figure shows that there is a temperature jump associated with the edge of the cavity at $r\sim40$\kpc , but it does not indicate a shock front. Instead it shows a temperature jump going from a cooler to warmer region (from $kT\sim2.8\keV$ to $kT\sim5.0\keV$), {\rm i.e.} it could be a cold front. In Section 5.2, we discuss in greater detail this front.     
\begin{figure}
\centering
\begin{minipage}[c]{0.98\linewidth}
\centering \includegraphics[width=\linewidth]{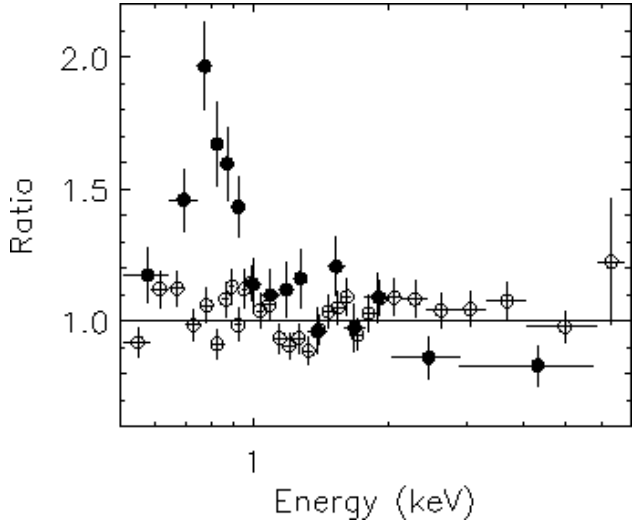}
\end{minipage}
\caption{Ratio of the spectrum from the region containing the plume-like feature (filled points) to the best-fitting {\sc mekal} model of the spectrum not containing the plume feature (non-filled points), both within $r=1-9''$. The normalization of the region containing the plume was determined by fitting the spectrum excluding data between $0.7-1$ keV. This plot clearly shows the excess of the Fe L emission feature associated with the plume.}
\label{fig11}
\end{figure}

\subsubsection{Plume-like feature}
According to the middle panel of Fig. \ref{fig6}, there is an increase in metallicity associated with the plume-like feature. We can also see from the left panel of Fig. \ref{fig6}, that there is an increase of temperature associated with the edge of the plume-like feature. To be more precise, in Fig. \ref{fig10}, we plot the temperature profile, electron density and pressure, as well as the count rate per pixel$^2$ (equivalent to the surface brightness distribution) as derived by selecting regions along the plume-like feature. However, we derive a count rate per pixel$^2$ for every arcsec in order to better trace the potential disturbances in the surface brightness profile. This figure shows a temperature jump at a radius similar to the one associated with the southern cavity ($r\sim40$\kpc). We associate this temperature jump with the same cold front that can be seen at the edge of the southern cavity, and which we discuss in Section 5.2. 

We now compare the spectrum of the plume to the one of a region of same radius ($r=1-9''$), but excluding the plume. The region containing the plume is taken as being between $290^o$ and $340^o$ (counter-clockwise from the north), and the remaining region is taken as the one excluding the plume ({\rm i.e.} a C-shape). In Fig. \ref{fig11}, we plot the ratio between the spectrum of the region containing the plume-like feature (filled points) to the best-fitting {\sc mekal} model of the spectrum not containing the plume feature (non-filled points). We used $\chi^2$ statistics and the normalization of the region containing the plume is determined by fitting the spectrum excluding data between $0.7-1$ keV. The excess of the Fe L emission feature associated with the plume is evident. 

This excess could nonetheless be caused by the colder temperature material associated with the cold front (and which coincides with the plume-like feature). However, even if we let the temperature free to vary, the ratio between the spectrum of the region containing the plume-like feature (filled points) to the best-fitting {\sc mekal} model of the spectrum not containing the plume feature still shows an excess in the Fe L emission feature. The excess is therefore still likely to be caused by a metallicity effect. 

Its nature could be due to iron enrichment by SNe, as suggested by \citet{Iwa2001328}. We can test this hypothesis by using the variable abundance model {\sc vmekal} in {\sc xspec}. Using the deep 100 ks $Chandra$ data, we analyse the spectrum of the plume-like feature within $r=1-9''$ (see Fig. \ref{fig8}). The best-fitting {\sc mekal} model finds a temperature of $kT=2.93\pm{0.11}$ keV and metal enrichment of $Z=0.74\pm{0.12}$ solar (equivalent to $Z_{\rm Fe}$ as the spectrum is dominated by iron emission lines). We also find a reduced $\chi^2$ statistic of 1.26.

If we test the {\sc vmekal} model using the enrichment pattern by SNIa, following the abundance ratios given by \citet{Dup2001548}, we find $kT=2.99\pm{0.11}$ keV, $Z_{\rm Fe}=0.47\pm{0.07}$ solar and a reduced $\chi^2$ statistic of 1.37 (same number of degrees of freedom). On the other hand, using the enrichment pattern by SNII, also following the abundance ratios given by \citet{Dup2001548}, we find $kT=2.90\pm{0.14}$ keV, $Z_{\rm Fe}=0.25\pm{0.06}$ solar and a reduced $\chi^2$ statistic of 2.02. Hence, the enrichment pattern by SNIa provides a much better fit, but not quite as much as the Solar photosphere enrichment pattern. We discuss these results in Section 5.3.

\section{Discussion}

\subsection{X-ray cavities}

There are at least two, and possibly three, X-ray cavities associated with the central AGN. The energy stored within each of the cavities can be estimated using Eq. \ref{eq1} \citep[{\rm e.g.} ][]{Bir2004607,Dun2005364,Raf2006652,Dun2006373,Dun2008385}. Here, $P$ is the thermal pressure of the ICM at the radius of the bubble and estimated from the X-ray data, $V$ is the volume of the cavity and for a relativistic fluid $\gamma_1$ = 4/3, therefore $E_{\rm bubble}=4PV$ \citep[this is also supported observationally, see][]{Gra2009386}. 
\begin{equation}
E_{\rm bubble}=\frac{\gamma_1}{\gamma_1-1}PV
\label{eq1}
\end{equation}

We assume that the cavities are of ellipsoidal shape. The volume is then given by $V=4{\pi}R^2_{\rm w}R_{\rm l}/3$, where $R_{\rm l}$ is the semi-major axis along the direction of the jet, and $R_{\rm w}$ is the semi-major axis perpendicular to the direction of the jet. 

The power injected into the medium is determined by dividing the energy of the bubble with its age. The latter is given by the buoyant rise time, the refill time or the sound crossing time. See respectively Eq. \ref{eq2} \citep[][]{Chu2001554}, Eq. \ref{eq3} \citep[][]{McN2000534} and Eq. \ref{eq4} \citep[][]{Mcn200745}. Here, $R$ is the distance from the radio point source to the middle of the cavity (projected), $S$ is the cross-sectional area of the bubble ($S={\pi}R_{\rm w}^2$), $C_{\rm D}$=0.75 is the drag coefficient \citep{Chu2001554}, $g$ is the local gravitational acceleration ($g=GM(<R)/R^2$), $r$ is the bubble radius ($r=(R_{\rm l}R_{\rm w})^{1/2}$ for an ellipsoidal bubble), and $c_\mathrm{s}$ is the sound crossing time ($c_\mathrm{s}=\sqrt{{\gamma_2}kT/(\mum_\mathrm{H})}$, where $kT$ is the plasma temperature at the radius of the bubble, $\gamma_2=5/3$ and $\mu=0.62$). 

\begin{equation}
t_\mathrm{buoyant}= R\sqrt{\frac{SC_{\rm D}}{2gV}}
\label{eq2}
\end{equation}
\begin{equation}
t_\mathrm{refill}= 2\sqrt{\frac{r}{g}}
\label{eq3}
\end{equation}
\begin{equation}
t_{c\mathrm{s}}=\frac{R}{c_\mathrm{s}}
\label{eq4}
\end{equation}

The buoyant rise time is the time it takes a bubble to reach its terminal buoyant velocity, which depends on the medium drag forces. This is a good estimate of a bubble's age that has clearly detached from their AGN and has risen, such as the cavity to the west in 4C+55.16. The refill time is the time it takes a bubble to rise buoyantly through its own diameter starting from rest (which is probably not the case). The sound crossing time is the time it takes a bubble to travel at the speed of sound. The latter is used under the assumption that the bubbles travel at subsonic speeds. Between the three time estimates, it is still not clear which is the best to use, although they should not vary significantly.  

\begin{table*}
\caption{Bubble properties \label{a2_t1a}}

\begin{tabular}{@{}lccccccccc@{}}
\hline
\hline
Cavity & $R_{\rm l}$ & $R_{\rm w}$ & $R_{\rm dist}$ & $t_{\rm sound}$ & $t_{\rm buoy}$ & $t_{\rm refill}$  & $P_{\rm sound}$ & $P_{\rm buoy}$ & $P_{\rm refill}$ \\
 & (kpc) & (kpc) & (kpc) & ($10^7$ yr) & ($10^7$ yr) & ($10^7$ yr) & ($10^{44}\ergps$) & ($10^{44}\ergps$) & ($10^{44}\ergps$) \\
\hline
south & 20.1 & 11.8 & 20.9 & 2.7 & 4.1 & 12.9 & 2.88 & 1.92 & 0.60 \\
north-west & 16.2 & 11.8 & 34.3 & 3.6 & 7.4 & 12.2 & 1.38 & 0.67 & 0.41  \\
west & 9.9 & 14.8 & 24.4 & 2.3 & 6.9 & 11.7 & 2.39 & 0.80 & 0.47  \\
\hline

\end{tabular}

\label{tab1}
\end{table*}

Table \ref{tab1} shows the estimates we obtain for the power being injected into the ICM by each of the bubbles. 4C+55.16 has a cooling luminosity in the $0.5-7\keV$ range of $L_{\rm cool}=(1.99\pm{0.02})\times10^{44}\ergps$ ($r_{\rm cool}=45$ kpc for a cooling time of 3 Gyr). The power being injected into the medium by all three cavities is therefore sufficient to prevent the gas from cooling, at least within a factor of two. The studies by \citet{Raf2006652} and \citet{Dun2005364} included 4C+55.16 in their sample of clusters with X-ray cavities. Our results are consistent with theirs, on average within a factor of two.

\subsection{Cold front}
Cold fronts in cool core clusters are contact discontinuities thought to originate from sloshing of low-entropy gas in the cluster core caused by disturbances on the central potential by phenomena such as past subcluster mergers \citep[see a review by ][]{Mar2007}. Simulations support this idea, but require that clusters with cold fronts have steep entropy profiles \citep[{\rm e.g.} ][]{Asc2006650}. This is generally the case for cool core clusters. Although cold fronts could harbour significant amounts of kinetic energy, it is not clear that they could dissipate it efficiently \citep{Mar2001562}. The exact cause of the sloshing is also not well known. Past mergers could cause sloshing of the cluster core \citep[as supported by simulations, see][]{Asc2006650}, but another possibility is whether outbursts from the central AGN could cause the core to recoil and slosh. 

In Section 4., we showed that 4C+55.16 has a clear temperature jump on the south-western side of the cluster associated with the southern cavity and plume-like feature, and that the plume-like feature is about twice as rich in metals than a region located within the same radius but excluding the plume (see the temperature and abundance map in Fig. \ref{fig6}). A more detailed analysis also showed that there is a discontinuity in the surface brightness profile associated with this front, see bottom panel of Fig. \ref{fig10} (although the break is at a radius slightly further away from the jump in temperature). These results are all consistent with the discontinuity being a cold front. 

In the literature, we find several examples of cold fronts showing a metallicity jump like in 4C+55.16, but with no clear discontinuity in the temperature profile, {\rm e.g.} A2052 \citep[][]{deP2010523}, A2199 and 2A 0335+096 \citep[][]{San2006371c}. Others do not show any evidence of a metallicity jump, but clearly show a discontinuity in temperature, such as A496 \citep{Dup2003583} or A2204 \citep{San2005356}. Our front also shows a break in entropy and pressure. The pressure remains constant across the front, a typical feature of cold fronts. Another key point that seems to support the idea that the front we observe is a cold front is the spiral (or tail-like) structure we see in Fig. \ref{fig3}. This figure was obtained by subtracting at every pixel, the average value of an ellipse passing through the point, centred on the X-ray point source. The figure shows an excess of emission ($\sim30$ per cent from the mean) in the shape of a large one-armed spiral structure extending from the metal-rich plume feature to the outskirts of the cluster. The mass within this structure is roughly $\sim6\times10^{11}M_{\odot}$, if we assume a cylindrical shape.  

Similar spiral-like structures have been seen in other clusters, such as in A2029 \citep{Cla2004616}, in 2A 0335+096 \citep{San2009396} and even in Perseus \citep[{\rm e.g. }][]{Chu2003590,Fab2000318}. Simulations in the context of cold fronts have shown that gaseous merging clusters with non zero impact parameters are able to reproduce these spiral structures as the central cool gas acquires angular momentum, in addition to the cold front \citep[{\rm e.g.} ][]{Asc2006650}. It takes several Gyrs for the cluster to regain a relaxed morphology. We expect 4C+55.16 to be relaxed, since the cluster shows no obvious signs of recent merger activity. The X-ray cavity to the south could therefore still have has the time to form, post merger, as bubble time scales are much shorter ($\sim0.1$ Gyr).  

If the discontinuities we see in temperature, abundance, pressure and entropy are not caused by a merger-induced cold front, they could simply by caused by feedback of the central AGN. As bubbles rise, they are expected to drag behind them, cool, metal-rich gas \citep{Chu2001554}. However, if there is originally cool low-entropy metal-rich gas at the base of the bubble, as its being inflated by the AGN, it could push the cooler metal-rich gas outwards. The gas would eventually slide back down around the bubble and would cause the appearance of a plume-like metal-rich feature running along the cavity, just like in 4C+55.16.

\subsection{Metal enrichment}
We now focus on analysing the metal enrichment associated with the plume-like structure. This structure is rich in the Fe emission feature (see Fig. \ref{fig11}). It is about $0.4Z_{\odot}$ more metal rich than a region located within the same radius, but excluding the plume (which has a metallicity of $\sim0.45Z_{\odot}$). If we assume that the plume has a cylindrical shape with a length of 30 kpc and width of 14 kpc, and that the density is $\sim0.065$ cm$^{-3}$ (based are the deprojected value along the plume), then the excess of iron mass is around $8.9\times10^6M_{\odot}$. It is worth mentioning that our results are based on the fits we have obtained using the abundance ratios of \citet{And198953}, whereas the more recent tables of \citet{Asp2009} or \citet{Lod2003} predict an iron abundance about 1.5 times lower. Refitting the data with these tables instead, we find that the plume has an iron excess of $\sim0.6Z_{\odot}$ instead of $\sim0.4Z_{\odot}$. However, the excess of iron in terms of mass does not change, as this quantity also depends on the relative abundance ratios, and the net effect cancels out. 

We can estimate the age of this structure in several ways. First, by using a measure of the diffusion coefficient, such as the one derived for Perseus in \citet[][ $2\times10^{29}\pcmsps$]{Reb2005359}, and an average size of the plume ($30\times14$ kpc$^2$), we find that time-scale associated with the plume should be on the order of $\sim6\times10^8$ yr. Second, we can estimate the age of the plume if we consider that it has to be younger than the local cooling time. The temperature and density of the plume then correspond to a cooling time of $\sim5\times10^8$ yr. We are ignoring heating here, which may be plausible since turbulence has not yet had the time to destroy it according to the long diffusion time. An age of about $5-6\times10^8$ yr implies that if the iron enrichment of the plume was caused by SNIa, then it would require around 2.5 SNIa per century, if we assume that 0.7$M_{\odot}$ are produced for every SNIa. This is at least an order of magnitude higher than the predicted rate of SNIa in an elliptical galaxy of similar $\rm\thinspace L_{B}$ \citep[see {\rm e.g.} ][]{Cap1997322}. 

The ages we derive for the plume are in agreement with those of the X-ray cavities ($t_{\rm bubble}\sim10^7-10^8$ yr, see Table 1), and are consistent with the idea that the plume could be created by the uprising of the southern bubble. Other examples of clusters have also shown evidence of significant quantities of metal-rich gas ($10^8-10^9M_{\odot}$) being uplifted by buoyantly rising bubbles. These include S{\'e}rsic 159-03 \citep{Wer2011}, M87 \citep{Wer2010407} and Hydra A \citep{Sim2009493}. In order for the plume-like structure to be long lived, magnetic fields could contribute to stabilizing it such as in the case of the emission-line filaments in Centaurus \citep{Tay2007382} and in Perseus \citep{Fab2008454}. 

In Section 4, we analysed the enrichment pattern of the plume using either SNIa or SNII enrichments. Past studies, including the previous one on 4C+55.16 by \citet{Iwa2001328}, suggested that the central iron excess seen in clusters was only caused by SNIa \citep[see also][]{Mat2003401}. However, our results led us to conclude that although the spectrum is more consistent with it being enriched by SNIa compared to SNII, the solar enrichment pattern still provides a better fit. \citet{San2006371b} also find that the enrichment pattern of the core of the Centaurus cluster is more consistent with Solar values, suggesting that both SNIa ($\sim70$ per cent) and SNII ($\sim30$ per cent) contribute to the enrichment \citep[see also ][]{deP2007465,Lov2011}. Our results agree with this picture.

\section{Concluding remarks}
4C+55.16 is X-ray bright ($\Lx\sim10^{45}\ergps$), radio powerful, and shows clear signs of interaction between its central galaxy and the ICM. 4C+55.16 has at least two, and possibly three X-ray cavities, two of which are filled with radio emitting particles. The power stored within them is around $6.7\times10^{44}\ergps$, and is sufficient to prevent the ICM from cooling. Our study confirms earlier results suggesting that there is a plume-like feature, running along one of the cavities, rich in the Fe L emission feature. However, we also find that the plume and cavity form a region of cool metal-rich gas that has a temperature and metallicity jump (by a factor of 2), and density jump. It could therefore be a cold front, or this could also be an example of metal enrichment by the central AGN, which has the potential to uplift cool metal-rich gas from the central galaxy. Finally, we find that the plume has an excess of iron ($M_{\rm Fe}=8.9\times10^6M_{\odot}$), and that its enrichment is more consistent with being Solar-like, and therefore suggesting that both SNIa and SNII contribute. This is in disagreement with the more general view that only SNIa enrich in the central regions of clusters. 

\section*{Acknowledgments}
JHL recognizes all the support given by the Cambridge Trusts, Natural Sciences and Engineering Research Council of Canada (NSERC), as well as the Fonds Quebecois de la Recherche sur la Nature et les Technologies (FQRNT). ACF thanks the Royal Society for support. GBT acknowledges support for this work from the National Aeronautics and Space Administration through $Chandra$ Award Number GO0-11139X.

\label{lastpage}
\bibliographystyle{mn2e}
\bibliography{bibli}
\end{document}